\documentstyle[12pt,epsf]{article}
\newcommand{\beqn}{\begin{eqnarray}}
\newcommand{\eeqn}{\end{eqnarray}}
\newcommand{\eq}[1]{(\ref{#1})}
\newcommand{\beq}{\begin{equation}}
\newcommand{\eeq}{\end{equation}}

\newcommand{\dual}{\mbox{}^{\ast}}

\makeatletter
\def\leqq{\mathrel{\mathpalette\gl@align<}}
\def\geqq{\mathrel{\mathpalette\gl@align>}}
\def\gl@align#1#2{\lower.6ex\vbox{\baselineskip\z@skip\lineskip\z@
    \ialign{$\m@th#1\hfil##\hfil$\crcr#2\crcr=\crcr}}}
\makeatother

\makeatletter
\def\sileqq{\mathrel{\mathpalette\gs@align<}}
\def\sigeqq{\mathrel{\mathpalette\gs@align>}}
\def\gs@align#1#2{\lower.6ex\vbox{\baselineskip\z@skip\lineskip\z@
    \ialign{$\m@th#1\hfil##\hfil$\crcr#2\crcr\sim\crcr}}}
\makeatother

\def\bbbone{{\mathchoice {\rm 1\mskip-4mu l} {\rm 1\mskip-4mu l}
{\rm 1\mskip-4.5mu l} {\rm 1\mskip-5mu l}}}
\def\bbbc{{\mathchoice {\setbox0=\hbox{$\displaystyle\rm C$}\hbox{\hbox
to0pt{\kern0.4\wd0\vrule height0.9\ht0\hss}\box0}}
{\setbox0=\hbox{$\textstyle\rm C$}\hbox{\hbox
to0pt{\kern0.4\wd0\vrule height0.9\ht0\hss}\box0}}
{\setbox0=\hbox{$\scriptstyle\rm C$}\hbox{\hbox
to0pt{\kern0.4\wd0\vrule height0.9\ht0\hss}\box0}}
{\setbox0=\hbox{$\scriptscriptstyle\rm C$}\hbox{\hbox
to0pt{\kern0.4\wd0\vrule height0.9\ht0\hss}\box0}}}}
\def\bbbe{{\mathchoice {\setbox0=\hbox{\smalletextfont e}\hbox{\raise
0.1\ht0\hbox to0pt{\kern0.4\wd0\vrule width0.3pt
height0.7\ht0\hss}\box0}}
{\setbox0=\hbox{\smalletextfont e}\hbox{\raise
0.1\ht0\hbox to0pt{\kern0.4\wd0\vrule width0.3pt
height0.7\ht0\hss}\box0}}
{\setbox0=\hbox{\smallescriptfont e}\hbox{\raise
0.1\ht0\hbox to0pt{\kern0.5\wd0\vrule width0.2pt
height0.7\ht0\hss}\box0}}
{\setbox0=\hbox{\smallescriptscriptfont e}\hbox{\raise
0.1\ht0\hbox to0pt{\kern0.4\wd0\vrule width0.2pt
height0.7\ht0\hss}\box0}}}}

\begin{document}

~
\vspace{-1cm}
\begin{flushright}
{\large ITEP-TH-36/98}

\vspace{0.2cm}
{\large KANAZAWA-98-08}

\vspace{0.2cm}
{\large UL--NTZ 18/98}

\vspace{0.3cm}
{\sl October 19, 1998}
\end{flushright}

\begin{center}

{\baselineskip=16pt
{\Large \bf {\boldmath$ Z$}--Vortex Percolation}\\
\vspace{0.3cm}
{\Large \bf in the Electroweak Crossover Region}

\vspace{1cm}

{\large
M.~N.~Chernodub\footnote{chernodub@vxitep.itep.ru}$^{\! ,a}$,
F.~V.~Gubarev\footnote{Fedor.Gubarev@itep.ru}$^{\! ,a}$,
E.--M.~Ilgenfritz\footnote{ilgenfri@hep.s.kanazawa-u.ac.jp}$^{\! ,b}$\\
and A.~Schiller\footnote{schiller@tph204.physik.uni-leipzig.de}$^{\! ,
c}$}\\

\vspace{.4cm}
{ \it

$^a$ ITEP, B.Cheremushkinskaya 25, Moscow, 117259, Russia

\vspace{0.3cm}

$^b$ Institute for Theoretical Physics, Kanazawa University,\\
Kanazawa 920-1192, Japan

\vspace{0.3cm}

$^c$ Institut f\"ur Theoretische Physik and NTZ, Universit\"at
Leipzig,\\ D-04109 Leipzig, Germany

}}
\end{center}
\vspace{4mm}
\abstract{
We study the statistical properties of $Z$--vortices and Nambu
monopoles in the $3D$ $SU(2)$ Higgs model for a Higgs mass $M_H
\approx 100$ GeV
near and above the crossover
temperature, where these defects are thermally excited.
Although there is no phase
transition at that strong selfcoupling, we observe that the
$Z$--vortices exhibit the {\it percolation} transition that has been
found recently to accompany the first order thermal transition that
exists at smaller Higgs mass.
Above the crossover temperature
percolating
networks of $Z$-vortex lines are ubiquitous, whereas vortices
form
a dilute gas of closed vortex loops and
(Nambu) monopolium
states on the low-temperature side of the crossover.  The percolation
temperature turns out to be roughly independent of the lattice spacing.
We find that the Higgs modulus is smaller (the gauge action is larger)
inside the vortices, compared to the bulk average.
This correlation becomes very strong on the low-temperature side.
The percolation transition is a prerequisite of some
string mediated baryon number generation scenarios.}

\baselineskip=14pt
%%% For alphabetic footnotes indices in text  %%%%
\setcounter{footnote}{0}
\renewcommand{\thefootnote}{\alph{footnote}}
%%%%%%%%%%%%%%%%%%%%%%%%%%%%%%%%%%%%%%%%%%%%%%%%%%

\section{Introduction}

According to recent lattice
studies~\cite{wirNP97,wirPRD97,LatticeCrossover,ThisDayReview} it is very
likely that the standard electroweak theory does not go through
a true phase transition at finite temperature.
The $3D$ $SU(2)$ Higgs model, which with a $t$-quark of appropriate mass is the
effective, dimensionally reduced version of the electroweak theory,
ceases to possess a first order transition for a
Higgs mass $M_H > 72$ GeV.  At larger values of Higgs mass the model
investigated merely experiences a smooth crossover~\cite{kajantieprl}.
Due to the quantitative similarity of the phase transitions in the
$SU(2)$ Higgs and in the $SU(2) \times U(1)$ Higgs~\cite{SU2U1} models
and taking the current experimental lower bound~\cite{HiggsMassLimit} of the
Higgs boson mass, $M_H \ge 89.3$ GeV,
into account, the statement above is highly justified.

Within the most popular baryon number generation (BG)
scenarios~\cite{BAUScenarios} a {\it sufficiently strong} first order transition
is required. Therefore the search for viable extensions of the standard model and
the study of their phase structure has become an
important direction of research.  Nevertheless, last but not least in view of
possible alternative BG mechanisms, it is still of some phenomenological
interest to study features of the {\it standard model} which change qualitatively
at some characteristic temperature (at the $W$ mass scale).
For instance, a variation of the diffusion rate of the Chern--Simons
number~\cite{Davis98} and a changing spectrum of screening
states~\cite{IlSchSt98} could be the analogue of the phase transition that
exists at low Higgs mass.

Very recently, we have shown \cite{ChGuIlSch98} that the first order
phase transition at $M_H \le 70$ GeV is accompanied by a percolation
transition\footnote{This is in agreement with an observation of
Ref.~\cite{antunes}, made in a different context, that a percolation
of strings is a good disorder parameter for a phase transition in
field theory.} experienced by certain kinds of topological defects. This
issue is a very new one in the field of lattice studies of
the electroweak transition, and various aspects have still to be
clarified. In this paper we shall describe what remains of this
percolation transition when the thermal phase transition changes into a
rapid crossover at higher Higgs mass.

This paper is the second of a series of studies we want to devote to the
{\it statistical properties}
of so-called {\it embedded} topological defects~\cite{VaBa69,BaVaBu94}
within the $SU(2)$ Higgs model.
The embedded defects of interest are Nambu monopoles~\cite{Na77} and
$Z$--vortices~\cite{Na77,Ma83}.
Although not stable topologically, these defects are seen
occurring as the result of thermal fluctuations.
Here we are able to show that the percolation transition mentioned above
persists at higher Higgs mass.  We also provide first evidence that
$Z$--vortices are indeed characterized by inhomogeneities of
gauge invariant quantities (gauge field action and Higgs field modulus) that
should be expected because of
the appearance of corresponding vortex solutions in the continuum.

Embedded topological defects might be important agents in some
electroweak baryogenesis mechanisms. One of such scenarios is based on
the decay of an electroweak string network as the Universe cools
down. According to this mechanism, long electroweak strings should decay
into smaller, twisted and linked string loops which carry non-zero
baryon number. This could then explain the emergence of non-zero
baryon number in the
Universe~\cite{StringScenario,Va94}.
We have nothing to contribute to this mechanism as such nor to the
kinetics of such decaying network. In our paper we merely
study the properties
of the embedded strings as they pop up in thermal equilibrium
of the $SU(2)$ Higgs model.
Similar to what we have found earlier in the symmetric phase
(when it can be clearly separated by a first order transition at lower
Higgs mass) our results show the existence of a network of $Z$-vortices on the
high-temperature side of the crossover, with finite probability
of percolating (at $T > T_{\mathrm{perc}}$),
while only smaller clusters occur below $T_{\mathrm{perc}}$, however
much more frequently than at lower Higgs mass. Some of these clusters have
open ends occupied by Nambu monopoles.
This suggests a BG scenario without a first order phase transition
according to which a large vortex cluster might
decay into small vortex pieces at some temperature.

In this work, a vortex cluster is a collection of connected dual
links which carry non-zero vorticity (are occupied by vortex
trajectories).  We have used an active bond percolation algorithm
known from cluster algorithms for spin models~\cite{Cluster} in order
to label various disconnected $Z$--vortex clusters that coexist in a
configuration.

Our investigation is performed in the framework of dimensional
reduction which is expected to lend a reliable effective description
of the $4D$ $SU(2)$ Higgs theory for Higgs masses between $30$ and
$240$ GeV at temperatures of $O(100)$ GeV.  Due to their relatively
rich abundance on the low temperature side in the crossover region,
the physics of distinct vortex loops and monopolium states (thought
to be the remnants of the percolation cluster(s)) is sufficiently
interesting to be studied more in detail and within the $4D$
Euclidean approach. Equally interesting would be the space-time
structure of the vortex networks in the percolating phase.  In our
present $3D$ approach, we can get only some rough anticipation of
what is going on in $4D$:  the average $3D$ densities might be {\it
time-projected} densities (describing the vortex network above or the
small vortex clusters below the percolation temperature).  In the
$4D$ Euclidean version of the model the vortex lines we are studying
would sweep out $2$-dimensional world surfaces.  The clusters of
filamentary embedded vortex defects that we observe in the
dimensionally reduced variant are compactifications of these world
surfaces or intersections with time slices.  In the present paper, we
have for the first time studied the $3D$ cluster statistics (the
number and length distributions) of the configurations.

The structure of the paper is as follows. We review the effective,
dimensionally reduced formulation of the electroweak theory as a
$3D$ $SU(2)$ lattice Higgs model
in Section~2. In this
section, for the convenience of the reader, we also formulate the lattice
definitions of the (elementary and extended) embedded topological
defects proposed in Refs.~\cite{ChGuIl97,ChGuIlSch98}.  In Section~3
we present the numerical results on the density of Nambu monopoles
and vortices and on the percolation probability of the corresponding
$Z$--vortex lines in the crossover region. We show how the average
number of lattice clusters (formed by vortex lines) and the average
length ({\it i.e.} the number of dual links with non-zero vorticity)
of vortex clusters changes at the transition. We also give some first
results on gauge invariant signatures showing that the vortices
provide a filamentary inhomogeneous structure of the system.
Section~4 contains a short discussion of our results and conclusions.

\section{Nambu Monopoles and $\mathbf{Z}$--Vortices in the Lattice
$\mathbf{SU(2)}$ Higgs Model}

For the investigation of the thermal equilibrium properties of the
defects to be defined below
we used the lattice $3D$ $SU(2)$ Higgs model with the
following action:
\beqn
S &=& \beta_G \sum_p \Bigl(1 - \frac{1}{2} \mbox{\rm Tr} U_p \Bigr)
- \beta_H \sum_{x,\mu} \frac{1}{2} \mbox{\rm Tr}
(\Phi_x^+ U_{x, \mu} \Phi_{x + \hat\mu})
 + \sum_x \biggl( \rho_x^2 + \beta_R (\rho_x^2-1)^2 \biggr)
\nonumber
\eeqn
(the summation is taken over plaquettes $p$, sites $x$ and
links $l=\{x,\mu\}$). The action contains three parameters: the gauge
coupling $\beta_G$, the lattice Higgs self-coupling $\beta_R$ and
the hopping parameter $\beta_H$.  The gauge fields are represented by
unitary $2 \times 2$ link matrices $U_{x,\mu}$ and $U_p$ denotes
the $SU(2)$ plaquette matrix. In this action,
the Higgs field is parametrized as follows: $\Phi_x = \rho_x V_x $,
where $\rho_x^2= \frac12 \mbox{\rm Tr}(\Phi_x^+\Phi_x)$ is the Higgs
modulus squared, and $V_x$ an element of the group $SU(2)$.  Later
on, the $2 \times 2$ matrix-valued Higgs field $\Phi_x$ is replaced
by the more standard isospinor notation $\phi_x =
{(\Phi^{11}_x,\Phi^{21}_x)}^T$.

The lattice parameters are related to the couplings of the
$3D$ superrenormalizable $SU(2)$ Higgs model in the continuum,
$g_3$, $\lambda_3$ and
$m_3(\mu_3=g_3^2)$ as given {\it e.g.} in \cite{wirNP97}.
As in \cite{wirNP97} a parameter $M_H^*$ is used (approximately equal
to the zero temperature physical Higgs mass) to parametrize the Higgs
self-coupling as follows:  \beqn \beta_R=\frac{\lambda_3}{g_3^2}
\frac{\beta_H^2}{\beta_G} = \frac{1}{8}
{\left(\frac{M_H^*}{80\ {\mbox {GeV}}}\right)}^2
\frac{\beta_H^2}{\beta_G}\, .
\label{MH*}
\eeqn
Lattice coupling $\beta_G$ and continuum coupling $g^2_3$ are related by
\beqn
\beta_G = \frac{4}{a g^2_3}\, , \label{betaG}
\eeqn
with $a$ being the lattice spacing. We have studied the model at
different gauge couplings $\beta_G$ in order to qualitatively
understand the appearance on the lattice of embedded defects of some
characteristic physical size using lattices with different lattice
spacing such that, eventually, the continuum limit can be accessed.
This requires to define operators which count
extended defects of arbitrary size in lattice units.

Let us first recall the definition of the elementary topological
defects. The gauge invariant and quantized lattice
definition~\cite{ChGuIl97} of the
Nambu monopole is closely related to the definition in the continuum
theory~\cite{Na77}.
First we define a composite adjoint unit
vector field $n_x = n^a_x \, \sigma^a$, $n^a_x = - (\phi^+_x
\sigma^a \phi_x) \slash (\phi^+_x \phi_x )$
with $\sigma^a$ being the Pauli matrices. In the following construction,
the field $n_x$ plays a
role similar to the direction of the adjoint Higgs field in the
definition of the 't~Hooft--Polyakov monopole~\cite{tHPo74} in the
Georgi--Glashow model. Here it is used to define
the gauge invariant flux~${\bar \theta}_p$ carried by the plaquette $p$,
\beqn
{\bar \theta}_p (U,n) & = & \arg \Bigl( {\mathrm {Tr}}
\left[(\bbbone + n_x) V_{x,\mu} V_{x +\hat\mu,\nu}
V^+_{x + \hat\nu,\mu} V^+_{x,\nu} \right]\Bigr)\, ,
\label{AP}
\eeqn
in terms of projected links
\beqn
V_{x,\mu}(U,n) & = & U_{x,\mu} + n_x U_{x,\mu} n_{x + \hat\mu}\, .
\nonumber
\eeqn

In the unitary gauge, with $\phi_x={(0,\varphi)}^T$ and $n^a_x \equiv
\delta^{a3}$, the phases $\theta^u_l = \arg U^{11}_l$ behave as a
compact Abelian field with respect to the residual Abelian gauge
transformations $\Omega^{abel}_x = e^{i \sigma_3 \, \alpha_x}$,
$\alpha_x \in [0,2 \pi)$ which leave the unitary gauge condition
intact. Instead of in terms of the plaquettes $\theta_p$ of this Abelian field,
the magnetic charge of Nambu monopoles - the topological
defects of this Abelian field - can alternatively
be defined using a gauge invariant
construction~\cite{ChGuIl97}. The monopole charge $j_c$ carried
by the cube $c$ can be expressed in terms of the gauge invariant
fluxes \eq{AP} passing through the surface $\partial c$:
\beqn
j_c = - \frac{1}{2\pi} \sum_{p \in \partial c}
{\bar \theta}_p\, , \quad
{\bar \theta}_p = \left( \theta_p - 2 \pi m_p \right) \in [-\pi,\pi)\, .
\label{j_N}
\eeqn

The $Z$--string~\cite{Ma83,Na77} corresponds to the
Abrikosov--Nielsen--Olesen vortex solution~\cite{ANO}
embedded~\cite{VaBa69,BaVaBu94} into the electroweak
theory\footnote{Note that there are two independent vortex solutions in the
electroweak theory: $Z$--vortex and $W$-vortex, see Ref.~\cite{BaVaBu94}.
In the limit of zero Weinberg angle $\theta_W$ (which is considered in
the present paper) both solutions coincide up to a global gauge
transformation. If $\theta_W \neq 0$ our construction \eq{SigmaN} of
the $Z$--vortex should be properly modified and complemented by one of
the $W$--vortex. In fact, \eq{SigmaN} would then correspond to what
is called there a $W$--vortex solution.}.
The $Z$--vorticity number corresponding to the
plaquette $p=\{x,\mu\nu\}$ is defined~\cite{ChGuIl97} as follows:
\beqn
\sigma_p = \frac{1}{2\pi} \Bigl( \chi_p - {\bar \theta}_p \Bigr) \, \, ,
\label{SigmaN}
\eeqn
where ${\bar \theta}_p$ has been given in \eq{AP}, and
$\chi_{p} = \chi_{x,\mu\nu} =
\chi_{x,\mu} + \chi_{x +\hat\mu,\nu} - \chi_{x + \hat\nu,\mu} -
\chi_{x,\nu}$ is the plaquette variable
formed in terms of the Abelian links
\beqn
\chi_{x,\mu} =
\arg\left(\phi^+_x V_{x,\mu} \phi_{x + \hat\mu}\right) \, .
\nonumber
\eeqn
The $Z$--vortex is formed by links $l=\{x,\rho\}$ of the dual
lattice which are dual to those plaquettes $p=\{x,\mu\nu\}$ which carry
a non-zero vortex number~\eq{SigmaN}: $\dual \sigma_{x,\rho} =
\varepsilon_{\rho\mu\nu} \sigma_{x,\mu\nu} \slash 2$. One can show
that $Z$--vortices begin and end on the Nambu (anti-) monopoles:
$\sum^3_{\mu=1} (\dual \sigma_{x-\hat\mu,\mu} - \dual \sigma_{x,\mu})
= \dual j_x$.

In order to understand the behavior of the embedded defects
towards the continuum limit we studied also numerically
so-called {\it extended} topological objects on the lattice according to
Ref.~\cite{IvPoPo90}.  A similar  approach has been pursued in
Ref.~\cite{Laine98}, in which lattice vortices of extended size have
been studied in the non-compact version of the $3D$ Abelian Higgs
model. An extended monopole (vortex) of physical size $k~a$ is
defined on $k^3$ cubes ($k^2$ plaquettes, respectively). The charge
of monopoles $j_{c(k)}$ on bigger $k^3$ cubes $c(k)$ is constructed
analogously to that of the elementary monopole, eq.  \eq{j_N}, with
the elementary $1\times 1$ plaquettes in terms of $V_{x,\mu}$
replaced by $n \times n$ Wilson loops (extended plaquettes). In the
context of pure gauge theory, in the maximally Abelian gauge, this
construction is known under the name of type-I extended objects. For
the present model an alternative construction (type-II), obtained by
blocking elementary topological objects, fails to lead to a good
continuum description~\cite{ChGuIlSch98}.  A more detailed definition
of extended Nambu monopoles and $Z$--vortices can be found in
Ref.~\cite{ChGuIlSch98}.

\section{Defect Dynamics at the Crossover}
\subsection{Density and Percolation}

First, we study the behavior of {\it elementary} Nambu monopoles and
$Z$--vortices along a line in parameter space
passing through the continuous crossover region. Monte Carlo
simulations have been performed on cubic lattices of size $L^3=16^3$
at $\beta_G=12$ for self-couplings $\lambda_3$ corresponding to a
Higgs mass $M_H^* = 100$~GeV, see eq.~\eq{MH*}. In our simulations
we used the algorithms described in Ref.~\cite{wirNP97} which combine
Gaussian heat bath updates for the gauge and Higgs fields with
several reflections for the fields to reduce the autocorrelations. We
varied the parameter $\beta_H$ in order to traverse the region of the
crossover at given $M_H^*$ and $\beta_G$.

At this stage we were interested in the behavior of the lattice Nambu
monopole ($\rho_m$) and $Z$--vortex ($\rho_v$) densities and of the
percolation probability $C$ for the $Z$--vortex as functions of $\beta_H$.
For each lattice
configuration, the densities $\rho_m$ and $\rho_v$ are given by
\beqn
 \rho_m = \frac{1}{L^3} \sum\limits_c |j_c|\, ,
 \qquad
 \rho_v = \frac{1}{3 L^3} \sum\limits_p |\sigma_p|\, ,
 \nonumber
\eeqn
where $c$ and $p$ refer to elementary cubes and plaquettes; the
monopole charge $j_c$ and the $Z$--vorticity $\sigma_p$ are defined
in \eq{j_N} and \eq{SigmaN}, respectively.

The percolation probability of the system of vortex lines
$\dual \sigma$ (which typically can be decomposed into several
lattice clusters) is defined
as a limit of the following two-point function~\cite{PoPoYu91}:
\beqn
C & = & \lim_{r \to \infty} C(r)\, ,\label{percolation}\\
C(r) & = & {\left(\sum\limits_{x,y,i}
\delta_{x \in \dual \sigma^{(i)}} \,\delta_{y \in \dual \sigma^{(i)}}
\cdot \delta\Bigl(|x-y|-r\Bigr) \right)} \cdot {\left(
\sum\limits_{x,y} \delta\Bigl(|x-y|-r\Bigr) \right)}^{-1}\, ,
\nonumber
\eeqn
where the summation is taken over all points $x$, $y$ of the dual lattice
with fixed distance and over all clusters $\dual \sigma^{(i)}$ of links
carrying vorticity ($i$ labels distinct vortex clusters).
The Euclidean distance between two points $x$ and $y$ is denoted as $|x-y|$.
The notation $x \in \dual \sigma^{(i)}$ means that the vortex world line
cluster $\dual \sigma^{(i)}$ passes through the point $x$.
Clusters $\dual \sigma^{(i)}$ are called percolating clusters if they
contribute to the limit $C$.

Formula \eq{percolation} corresponds to the thermodynamical limit.
In our finite volume we find numerically that the function $C(r)$ can
be fitted as $C(r) = C + C_0 r^{-\alpha} e^{-m r}$, with $C$, $C_0$,
$\alpha$ and $m$ being fitting parameters. As we observed, $m \sim
a^{-1}$ in the
explored region of the phase diagram, therefore we can
be sure that finite size corrections to $C$ are exponentially
suppressed.
If $C$ does not vanish on an infinite lattice
then the vacuum is populated by
one or more percolating clusters, each consisting of
{\it infinitely many} dual links with non-vanishing vorticity.
This implies the existence
of a non-vanishing vortex condensate.
If $C$ turns to zero the vortex condensate
vanishes according to this definition.

In Figure~1(a) we show the ensemble averages of densities,
$\langle \rho_m \rangle$ of Nambu monopoles and $\langle \rho_v
\rangle$ of
$Z$--vortices as a function of the hopping parameter $\beta_H$
for Higgs mass $M_H^*= 100$~GeV at gauge coupling $\beta_G = 12$.
Both densities vanish very smoothly with increasing hopping
parameter $\beta_H$, (which corresponds to a decreasing physical
temperature). The percolation probability shown in Figure~1(b) vanishes
at some value of the coupling constant $\beta_H$ corresponding to a
percolation temperature well above the temperature
where the densities turn to zero. In fact, the percolation ends while
the density of monopoles and vortices still amounts to some
fraction of the densities of these objects deep in the
symmetric phase.

We interpret this by an analogy according to which
the would-be Higgs phase in the crossover region (at temperatures below
some crossover temperature) resembles a type II superconductor in so far as
it can support thick vortices which cannot form infinite clusters.
In Nature, in a real cooling process passing the crossover, the percolating
cluster(s) that has (have) existed above the crossover temperature
would have been
broken into vortex rings or vortex strings connecting Nambu monopole pairs.

It would be tempting to identify the crossover temperature
with the percolation temperature if the latter has a well-defined meaning
in the continuum limit.
In order to explore whether the percolation effect persists approaching
the continuum limit we have studied also extended topological objects (using
the so-called type-I construction mentioned in
Section~2). According to \eq{betaG} the physical size of the $k^3$
monopoles (or the $k^2$ vortices) in simulations done at $\beta_G =
k~\beta^{(0)}_G$ should be roughly the same for all $k$. Since
we expect finite volume effects to be potentially more severe at $M_H^*=100$ GeV
than at a strongly first order phase transition
we were careful to keep also the physical size of the lattice fixed: at
$\beta_G = k~\beta^{(0)}_G$ we have performed simulations on lattices with a
volume ${(k~L_0)}^3$, respectively.

For the coarsest lattice we have chosen $\beta_G=\beta^{(0)}_G=8$ and
$L=L_0=16$. We show the behavior of the percolation probability near
the crossover point on Figures~2(a,b,c) for $k=1,2,3$, respectively.
One can clearly see the existence of a percolation transition for
each vortex size $k$. The actual value of the coupling
$\beta^{\mathrm{perc}}_H$ corresponding to the percolation transition varies
with $k$, similar to how $\beta^{\mathrm{trans}}_H$ was observed to change
with $\beta_G$ and $L^3$ at smaller Higgs mass (when there is a true first
order phase transition).  Also here, one can analogously define a physical
percolation temperature $T^{\mathrm{perc}}$ corresponding to
$\beta^{\mathrm{perc}}_H$. This temperature is found to become roughly
independent of $\beta_G$ (or the lattice spacing) with decreasing
lattice spacing. That means that the defects popping up dynamically
(as well as the corresponding percolation temperature) become increasingly
well-defined when the lattice becomes fine-grained enough to
resolve the embedded vortices as extended objects.
This indicates that the percolation temperature has a good chance
to possess a well-defined continuum limit.
Taking into account the perturbative relations between
$3D$ and $4D$ quantities~\cite{generic},
$T^{\mathrm{perc}}$ can be roughly estimated as 170
GeV or 130 GeV, dependent on which version of the $4D$ continuum
$SU(2)$-Higgs theory is represented by the effective model,
without fermions or with fermions including the top quark.
The corresponding zero
temperature Higgs mass $M_H$ would be 94 and 103 GeV in the
respective theories.

Notice however that for the finer lattices (bigger vortex size in
lattice units) there appears a tail of small percolation probability
before $C$ finally turns to zero. This means that, still above the
percolation temperature, the percolating clusters become more
dilute, only a part of the lattice configurations actually
contains a percolating cluster (''intermittent'' percolation), and
more and more smaller clusters appear.

\subsection{Cluster Statistics}

In order to look closer for the properties of the small vortex loops
that populate the low temperature side of the crossover at not too
low temperature, we have
also measured the Monte Carlo ensemble average of the number of
clusters per configuration and the average length per cluster.
The behavior of these quantities for different $k=1,2,3$ are
qualitatively the same, therefore we present these quantities for the
case of extension parameter $k=2$. We show in Figure~3 results
obtained for the corresponding lattice size $32^3$ and gauge coupling
$\beta_G=16$:  (a) the density of the Nambu monopoles and
$Z$--vortices, (b) the average length ${\cal L}$ per
$Z$--vortex cluster and (c) the average number ${\cal N}$ of
$Z$--vortex clusters per lattice configuration.

It is clearly seen from Figure~3(a) that in the region of
the percolation
transition (compare Figure~2(b) for $k=2$ at $\beta^{\mathrm{perc}}_{H}
\approx 0.3432$) the density of monopoles and vortices
decreases smoothly with increasing $\beta_H$.
From Figure~3(b) one can conclude that the average length of the
vortex clusters ${\cal L}$ decreases drastically while the average
number of vortex clusters ${\cal N}$ increases sharply (Figure~3(c))
already at somewhat smaller $\beta_H$. The latter reaches its maximum
at $\beta^{\mathrm{perc}}_{H}$.

With the help of equation \eq{betaG} and $g_3^2\approx g_4^2 T \approx 0.43 T$
we get, choosing $k=2$ and $\beta_G=16$ as an example, a lattice spacing
$a=1/(300~\mathrm{GeV})$.
We can estimate the densities of $k=2$ Nambu monopoles and $Z$--vortices
in physical units.  At the percolation temperature $170$ GeV we get,
for lattice densities $\rho_v=0.14$ and $\rho_m=0.26$,
in the continuum a vortex density of $(55~\mathrm{GeV})^2$
and a monopole density of $(93~\mathrm{GeV})^3$.
Taking into account that, classically, the core widths of embedded defects
are of the order of $M_W^{-1}$ we conclude that vortices and monopoles are
densely packed at the percolation transition and, as a result, their cores
are strongly overlapping.
After the percolation transition is completed, the open or closed vortex
strings are very short:  their mean length is roughly three times
the classical vortex width.
The average cluster length amounts to $~5ka=1/(30~\mathrm{GeV})$.

Although these results characterize structures formed by thermal fluctuations
within equilibrium thermodynamics, all
facts suggest strongly that in a non-equi\-li\-bri\-um cooling process
the percolating cluster(s) and other bigger clusters decay
near the percolation temperature $T^{\mathrm{perc}}$,
primarily into small closed vortex loops which later gradually shrink
and disappear with further decreasing of temperature (increasing of $\beta_H$).
The detailed dynamics of this process requires thorough investigations
using non-equi\-li\-bri\-um techniques.

Note that even at $\beta_H$ values far above the percolation point
there is a long plateau in the average length per vortex clusters at
a level of ${\cal L} \approx 2$. A more detailed analysis of the
field configurations shows in fact that here, below the percolation
temperature, each configuration contains a large number (a few dozen,
according to Figure~3(c)) of monopole--anti-monopole pairs connected
by vortex trajectories of length $1-2$ and only few additional closed
vortex loops. As it was shown in Ref.~\cite{HiJa94,ChGuIl97} the
sphaleron configuration (an unstable solution to classical equations
of motion) contains in its center a monopole--anti-monopole pair
connected by a short vortex string. It is suggestive to interpret
at least some part\footnote{The vortex string inside the electroweak
sphaleron is shown to be twisted~\cite{Va94,HiJa94,twist}. In
our measurements we do not check the twist of the $Z$--vortex segments,
therefore we are not able to relate all open vortex loops to sphalerons
with confidence.}
of the open $Z$--vortex strings (which exist within some temperature
interval below the percolation temperature with a small density) as
sphalerons.

\subsection{$Z$--Vortices as Physical Objects}

In this Subsection we show that our construction of the $Z$--vortices
on the lattice \eq{SigmaN} defines objects resembling some characteristic
features of the classical $Z$-vortex solutions in continuum. Our construction
detects a line-like object with non-zero vorticity
and there is no guarantee that this configuration has a particular
vortex profile. However, as we show below, the lattice $Z$--vortices
have some common features with the classical solutions.

At the center of a classical continuum $Z$--vortex the Higgs field
modulus is zero and the energy density reaches its
maximum~\cite{Na77,BaVaBu94}. If the vacuum is populated by
vortex-like configurations then it would be natural to expect that
along the axis of these configurations there will be a line of zeroes of
the Higgs field and of points with maximal energy density. The simplest way to
test how good this expectation is fulfilled is to measure the (squared)
modulus of the Higgs field and the energy density near the dual
vortex-carrying links
defined by \eq{SigmaN} and compare these quantities
with the corresponding bulk average values far from the vortex
core\footnote{A similar method has been used to study the physical
features of the
Abelian magnetic monopoles in non-Abelian gauge theories in
Ref.~\cite{MIP}.}.

Here we restrict ourselves to the case of elementary defects.
This is the worst case in the sense that it
obviously does not allow to define a profile of a defect
resembling the continuum case. We will come back to the profile of
an extended defect in a forthcoming publication.
For the present purposes the non-vanishing of \eq{SigmaN} is simply
used as a trigger to measure the above mentioned observables.
We define the mean value of the modulus squared of the Higgs field
inside the (elementary) vortex, ${<\rho^2>}_{\mathrm{in}}$,
as the average of $(\phi^+_x \phi_x )$
over the corners of {\it all} plaquettes with $\sigma_P \ne 0$ (dual to the
vortex-carrying links).
For simplicity, the analogous quantity outside the (elementary) vortex,
${<\rho^2>}_{\mathrm{out}}$,
is obtained as the corresponding average over the corners of
{\it all} plaquettes with $\sigma_P = 0$. Even more straightforward
is the definition of the corresponding averages of the gauge field
energy as volume
averages of $~1-\frac12 \mathrm{Tr} U_p~$ depending on whether $\sigma_P$
is equal to or different from zero.

The quantities ${<\rho^2>}_{\mathrm{in,out}}$ are plotted {\it vs.}
$\beta_H$
for $\beta_G=8$ in Fig.~\ref{corr}(a). One can clearly see that
the modulus of the Higgs field is smaller near the vortex trajectory
than outside the vortex for all values of the coupling
$\beta_H$.  In order to make clearer the different behavior of the quantities
${<\rho^2>}_{\mathrm{in,out}}$ on both sides of the crossover
we show in Fig.~\ref{corr}(b) the histograms of
these quantities on the symmetric side
($\beta_H=0.348$) and the Higgs side ($\beta_H=0.356$).

On the symmetric side of the crossover (smaller $\beta_H$) the
difference\footnote{Note, that only the difference between quantum
averages of the squared modulus of the Higgs field (not the quantum average
itself!) may be related to the continuum limit due to an additive
renormalization, see Ref.~\cite{Laine98} for a detailed discussion on
this point.} between quantities ${<\rho^2>}_{\mathrm{in,out}}$ is much
smaller
than on the Higgs side
(larger $\beta_H$).  This fact is natural, since on the symmetric side
the vortices densely populate the vacuum and vortex cores are
overlapping while on the Higgs side of the transition the vortices are
dilute. The value of the Higgs field modulus in the region between
closely placed vortices is smaller than
between far separated vortices.
As to be expected for elementary lattice vortices,
${<\rho^2>}_{\mathrm{in}}$ does not
vanish. This is due to the relatively large lattice spacing which prevents
that the Higgs field can be measured arbitrarily near the vortex axis.
A detailed study of the vortex profile (and a localization of the Higgs zeroes)
requires the extended vortex construction and is under
investigation~\cite{InPreparation}.

The gauge field energies ${<1-U_P>}_{\mathrm{in,out}}$ are plotted {\it
vs.}
$\beta_H$ in Fig.~\ref{corr}(c), the corresponding histograms are
shown in Fig.~\ref{corr}(d) for the symmetric side ($\beta_H=0.348$)
and the Higgs side ($\beta_H=0.356$) of the crossover. Both figures
show that the value of the gauge field energy near the vortex center
is larger than in the bulk on both sides of the crossover.

The results of this Subsection clearly show that $Z$--vortices are
physical objects: vortices carry excess gauge field energy and the
Higgs modulus decreases near the vortex center. Whether these
thermal excitations really resemble the features of the classical
continuum vortex solutions remains to be seen extending this study to
finer lattices.

\section{Discussion and Conclusions}

Recently we have started to investigate numerically the behavior of
Nambu mo\-no\-po\-les and $Z$--vortices in the $SU(2)$ Higgs model at high
temperatures within the dimensional reduction approach.
This model is used as representative for the standard model.
Here we have extended our previous work to a region of Higgs mass where
the model is known not to have a true thermal phase transition.
We show that the 3D percolation
transition that we have recently found to be a companion of the
first order phase transition at low Higgs mass, still exists at
the large but not unrealistic Higgs mass of $M_H\approx 100$ GeV.
At temperatures
above $T^{\mathrm{perc}} \approx 170$ GeV ($ \approx 130$ GeV
for the more realistic case with the top quark included) space is
densely populated by large vortex clusters with one or few infinitely
extended ones among them.
This state is not thermodynamically relevant at lower temperature.
Instead of very large clusters, a gas
of closed vortex loops and monopolium bound states (Nambu
monopole--anti-monopole pairs bound by $Z$--strings) prevails. Further
below $T^{\mathrm{perc}}$, with decreasing temperature small vortex loops
shrink and disappear while monopole--anti-monopole pairs still survive.
In the spirit of our recent investigation of classical sphaleron
configurations we would like to associate at least some part of these pairs
with sphaleron-like configurations known to exist in the broken phase.
Without going into details of specific mechanisms, we want to point
out that the non-equi\-lib\-ri\-um break-up of infinitely extended
electroweak vortex clusters into small closed loops with decreasing
temperature is a prerequisite of some string--mediated baryon number
generation scenarios \cite{StringScenario,Va94}. It is interesting to
see that similar defect structures are realized within our effective
Higgs model.

\section*{Acknowledgments}

M.~N.~Ch. is grateful to L.~McLerran, M.~I.~Polikarpov
and K.~Rummukainen for interesting discussions.

M.~N.~Ch. and F.~V.~G. were partially supported by the grants
INTAS-96-370, INTAS-RFBR-95-0681, RFBR-96-02-17230a and
RFBR-96-15-96740.

\newpage

\section*{Figures}

\begin{figure*}[!tbh]
\begin{center}
\begin{tabular}{cc}
\hspace{-0.8cm}\epsfxsize=7.1cm\epsffile{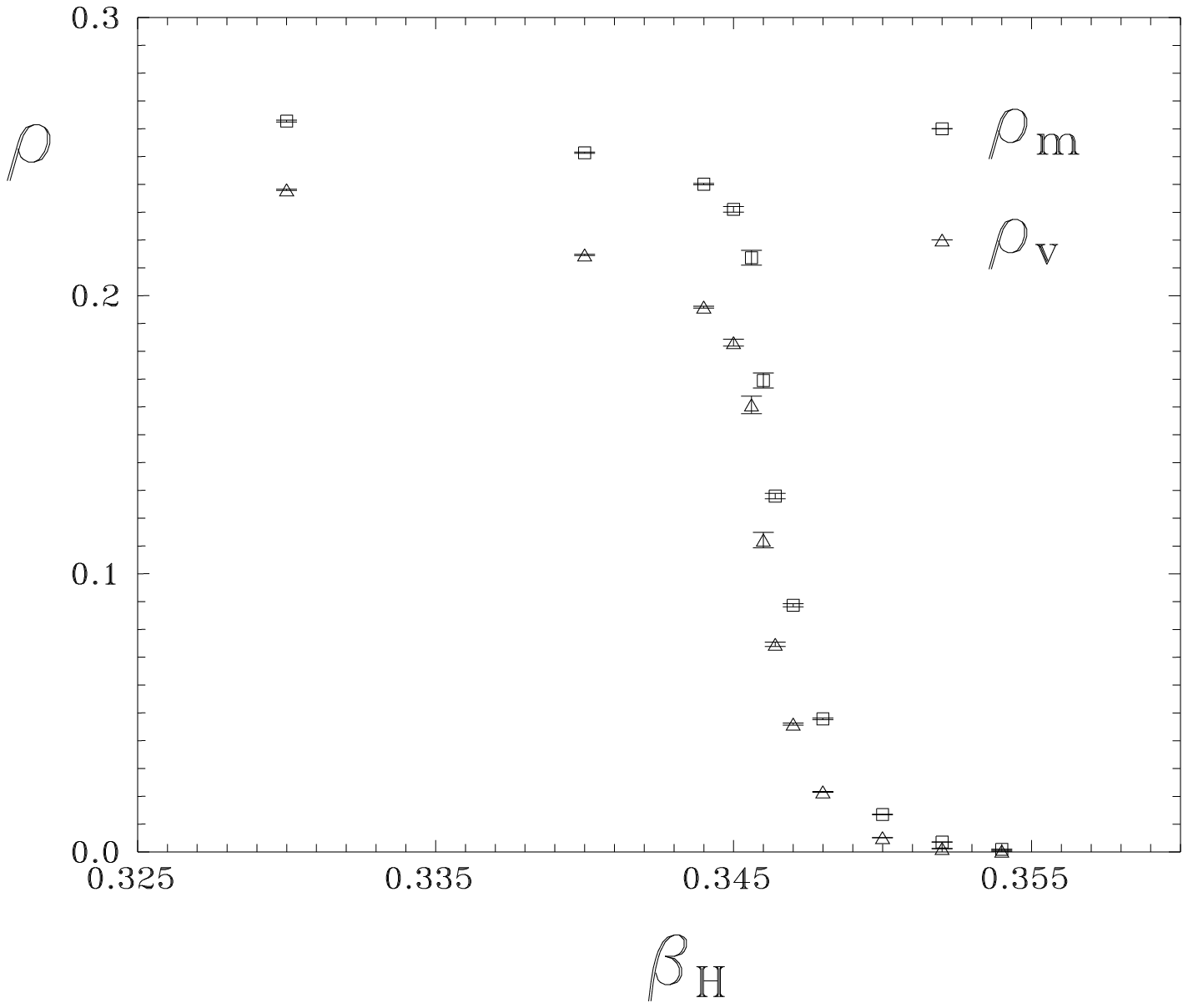} &
\hspace{0.8cm}\epsfxsize=7.1cm\epsffile{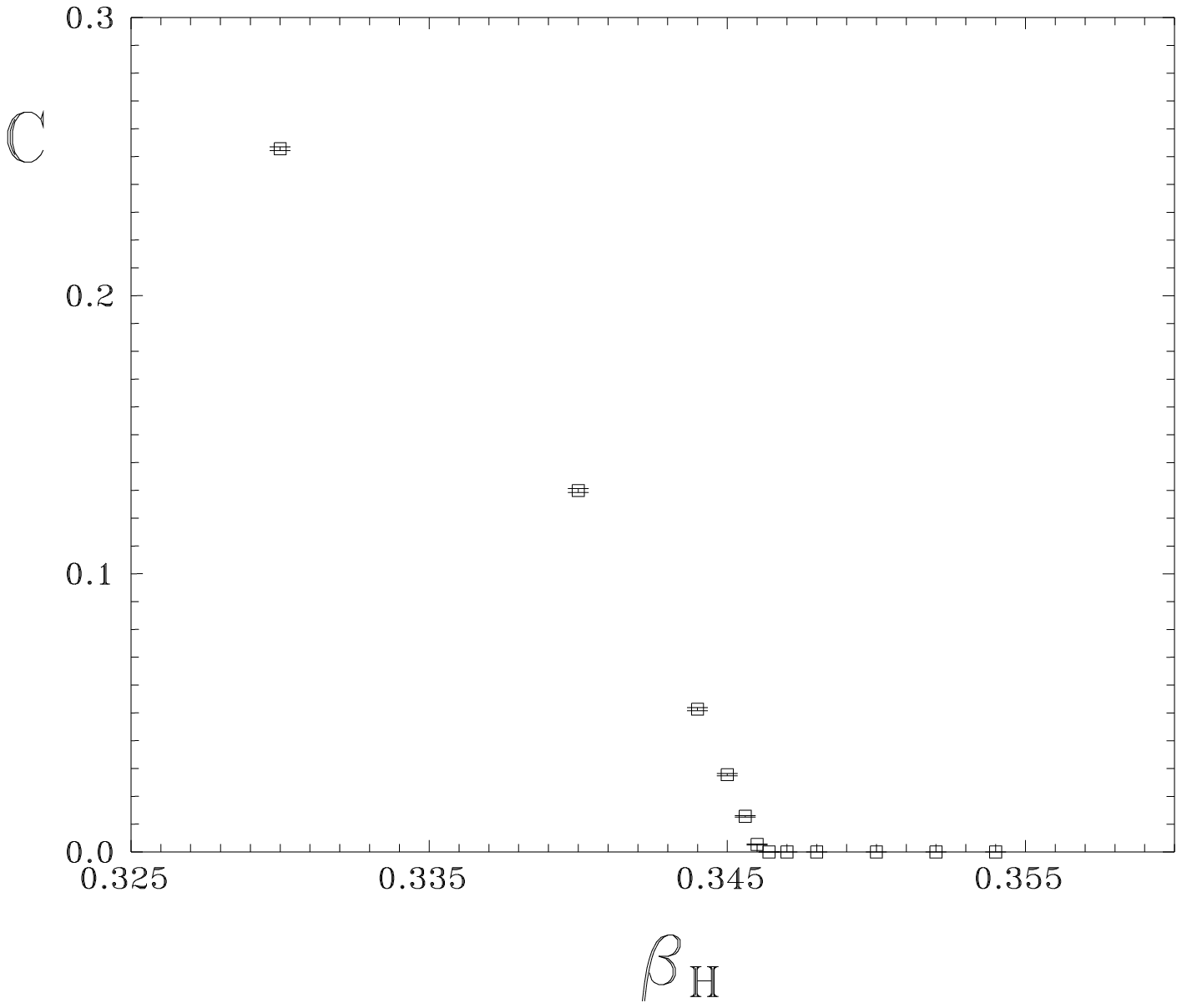} \\
(a) & \hspace{1.5cm}  (b) \\
\end{tabular}
\end{center}
\vspace{-0.5cm}
\caption{(a) Density $\rho_m$ of Nambu monopoles and $\rho_v$
of $Z$--vortices {\it vs.} hopping parameter
$\beta_H$ for Higgs mass $M_H^*= 100$~GeV at
gauge coupling $\beta_G = 12$; (b) Percolation probability $C$ of
$Z$--vortex trajectories for the same parameters;
the percolation transition happens at critical $\beta^{\mathrm{perc}}_H \approx
0.3451$.}
\end{figure*}

\begin{figure*}[!tbh]
\begin{tabular}{cc}
\hspace{-0.8cm}\epsfxsize=7.2cm\epsffile{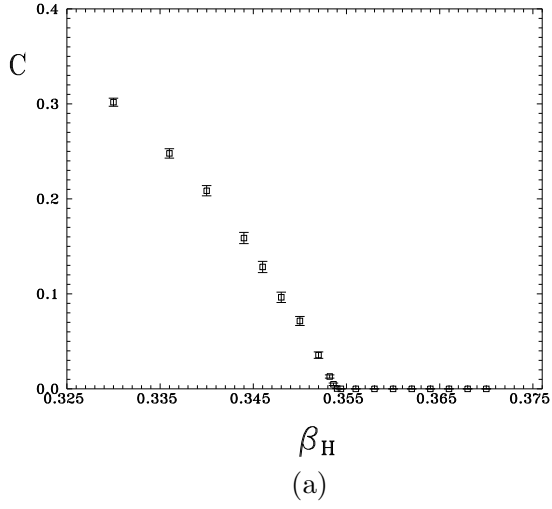} &
\hspace{0.8cm}\epsfxsize=7.1cm\epsffile{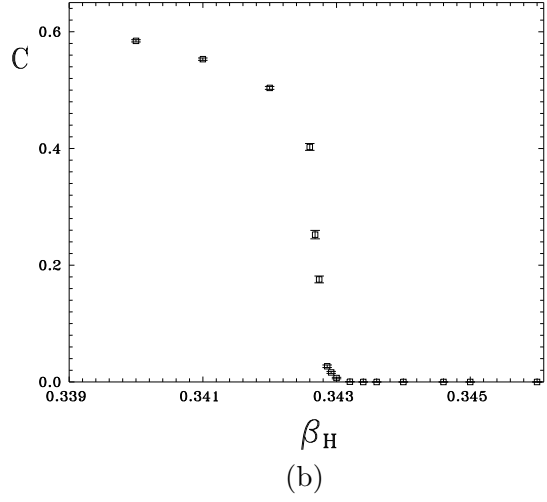} \\
(a) & \hspace{1.5cm}  (b) \vspace{1cm}\\
\end{tabular}
\centerline{\epsfxsize=7.2cm\epsffile{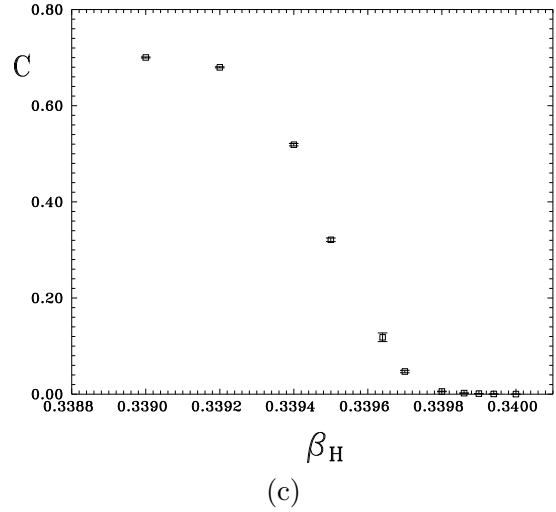}}
\centerline{(c)}
\vspace{-0.5cm}
\caption{Percolation probability $C$ {\it vs.} $\beta_H$ for
$Z$--vortices of size $k^2$ on lattices ${(16 \, k)}^3$ at
gauge couplings $\beta_G=8 \, k$, respectively, for
(a) $k=1$, (b) $k=2$ and (c) $k=3$.}
\end{figure*}

\begin{figure*}[!tbh]
\begin{tabular}{cc}
\hspace{-0.8cm}\epsfxsize=7.1cm\epsffile{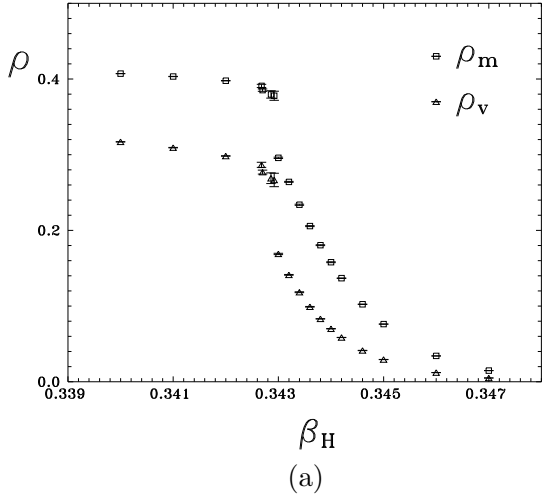} &
\hspace{0.8cm}\epsfxsize=7.1cm\epsffile{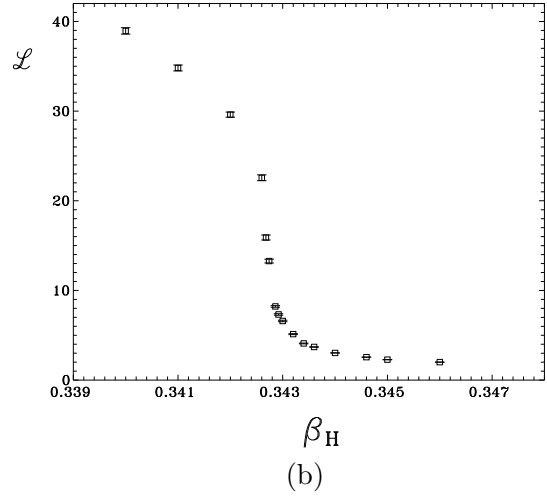} \\
(a) & \hspace{1.5cm}  (b) \vspace{1cm}\\
\end{tabular}
\centerline{\epsfxsize=7.1cm\epsffile{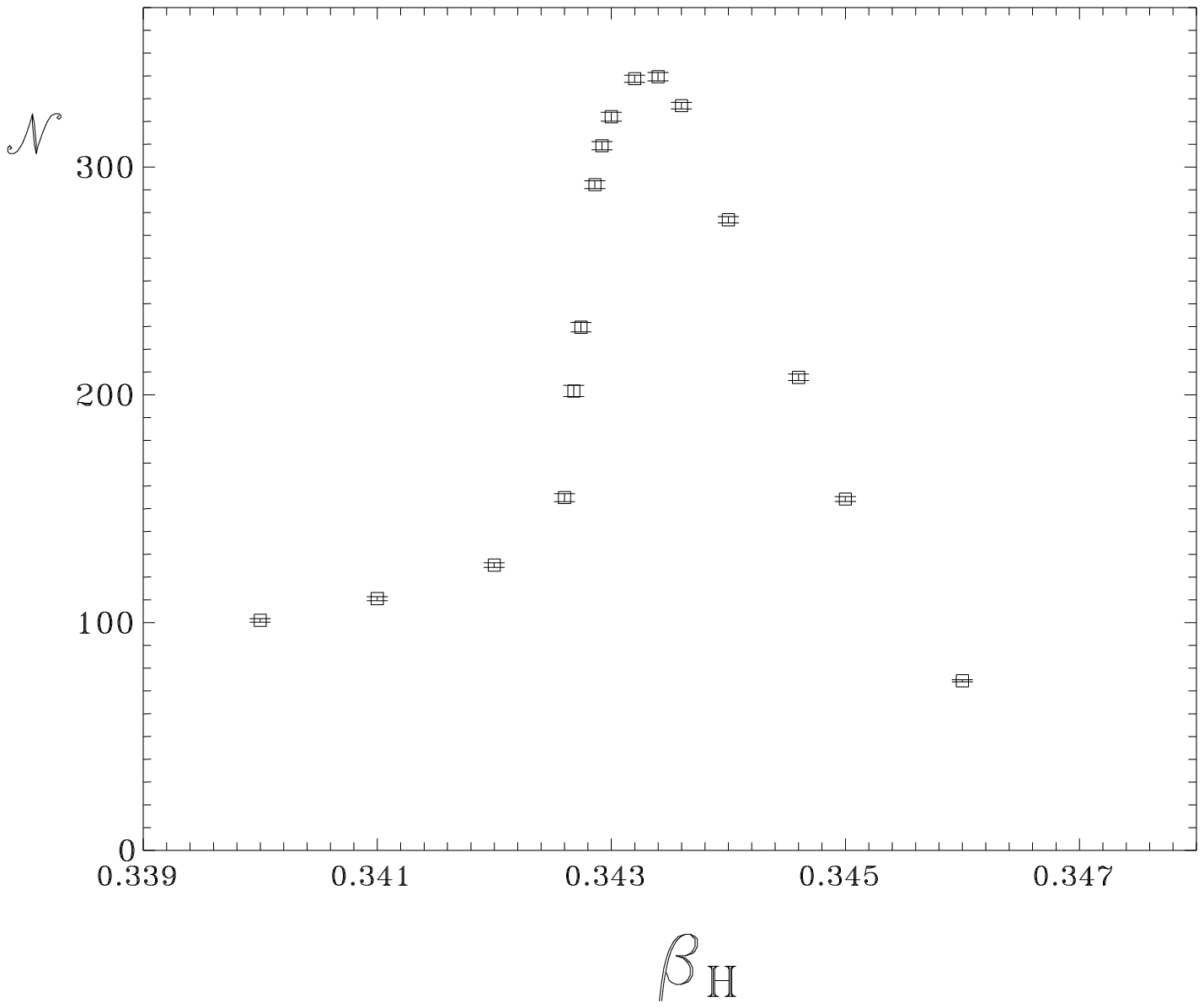}}
\centerline{(c)}
\vspace{-0.5cm}
\caption{(a) Density $\rho_m$ of Nambu monopoles and $\rho_v$
of $Z$--vortices,
(b) average length $\cal L$ per $Z$--vortex cluster and (c)
average
number ${\cal N}$ of
$Z$--vortex clusters per lattice
configuration (all for size $k=2$ objects with corresponding lattice
size $32^3$ and gauge coupling $\beta_G=16$).}
\end{figure*}

\begin{figure*}[!tbh]
\begin{center}
\begin{tabular}{cc}
\hspace{-0.8cm}\epsfxsize=7.2cm\epsffile{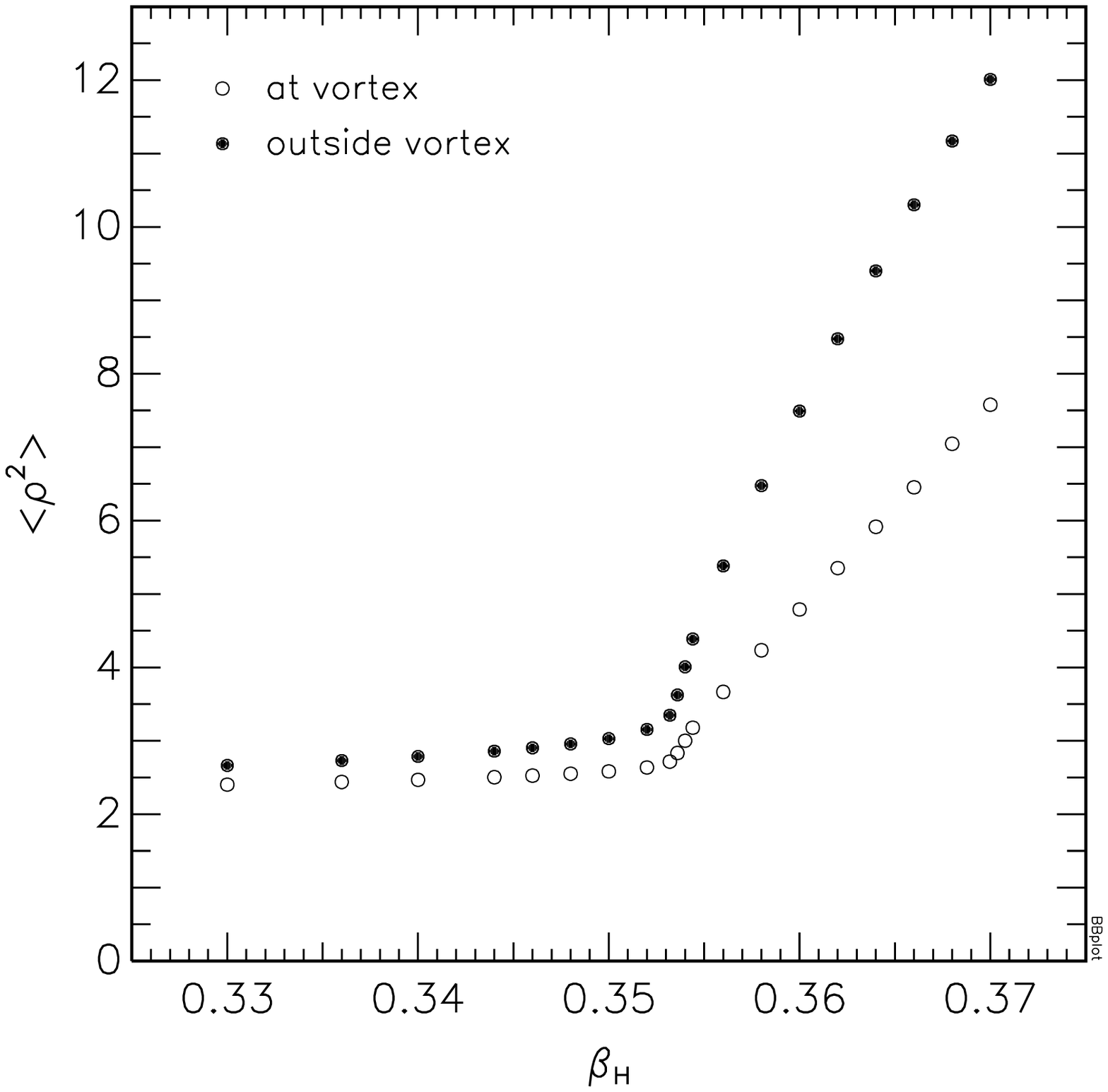} &
\hspace{0.8cm}\epsfxsize=6.75cm\epsffile{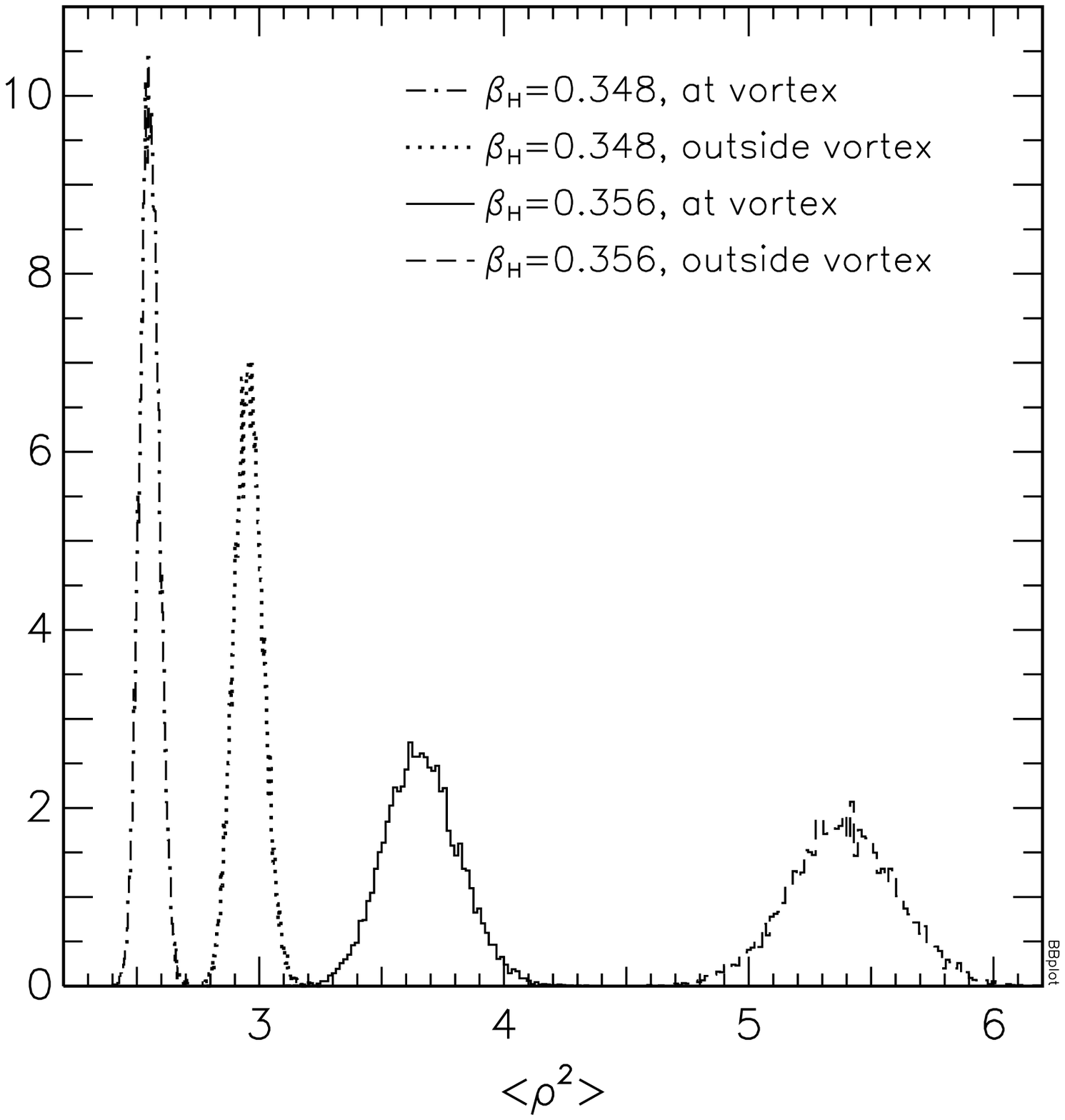} \\
(a) & \hspace{1.5cm}  (b) \vspace{0.5cm}\\
\end{tabular}
\begin{tabular}{cc}
\hspace{-0.8cm}\epsfxsize=7.2cm\epsffile{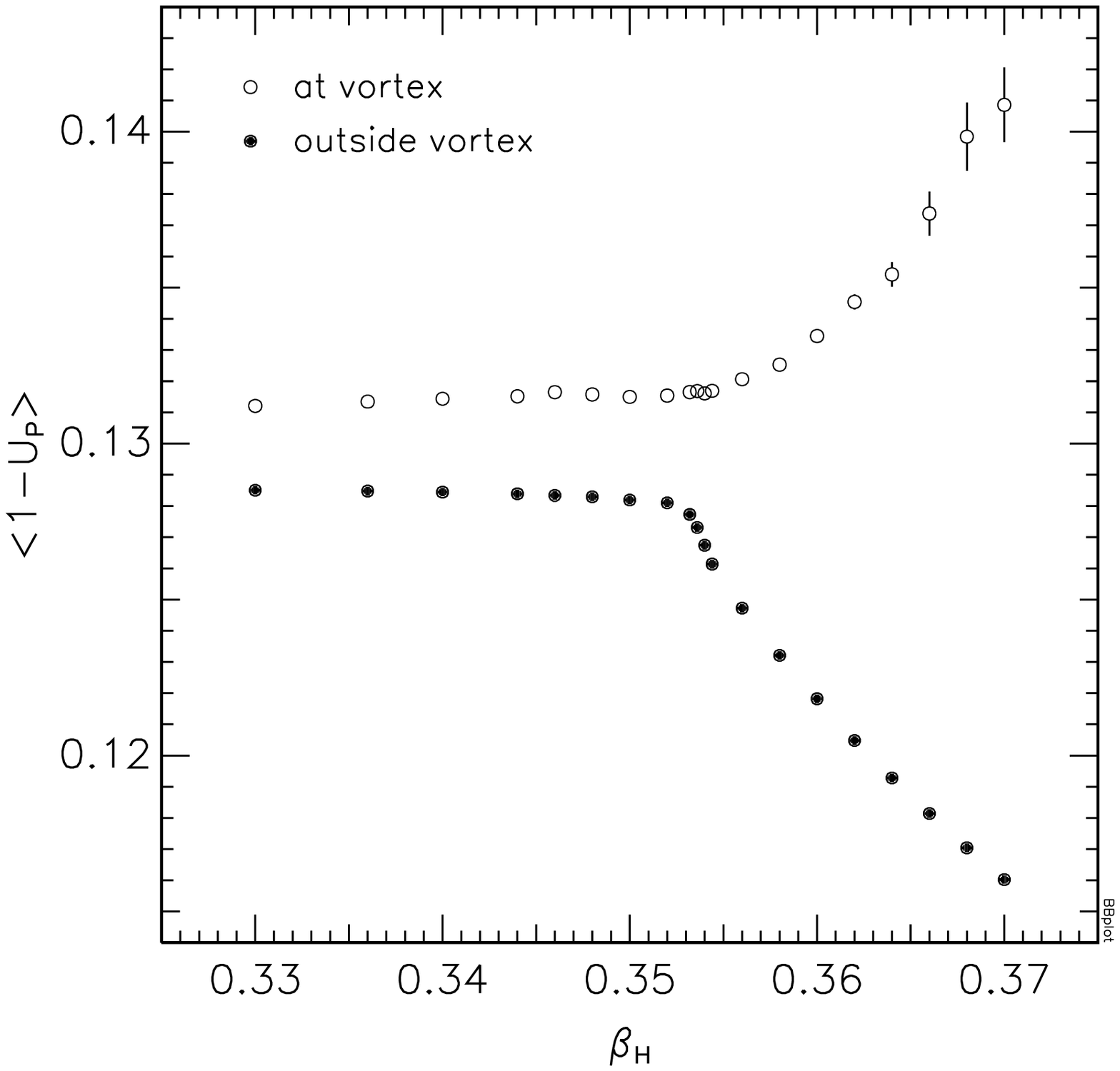} &
\hspace{0.8cm}\epsfxsize=6.75cm\epsffile{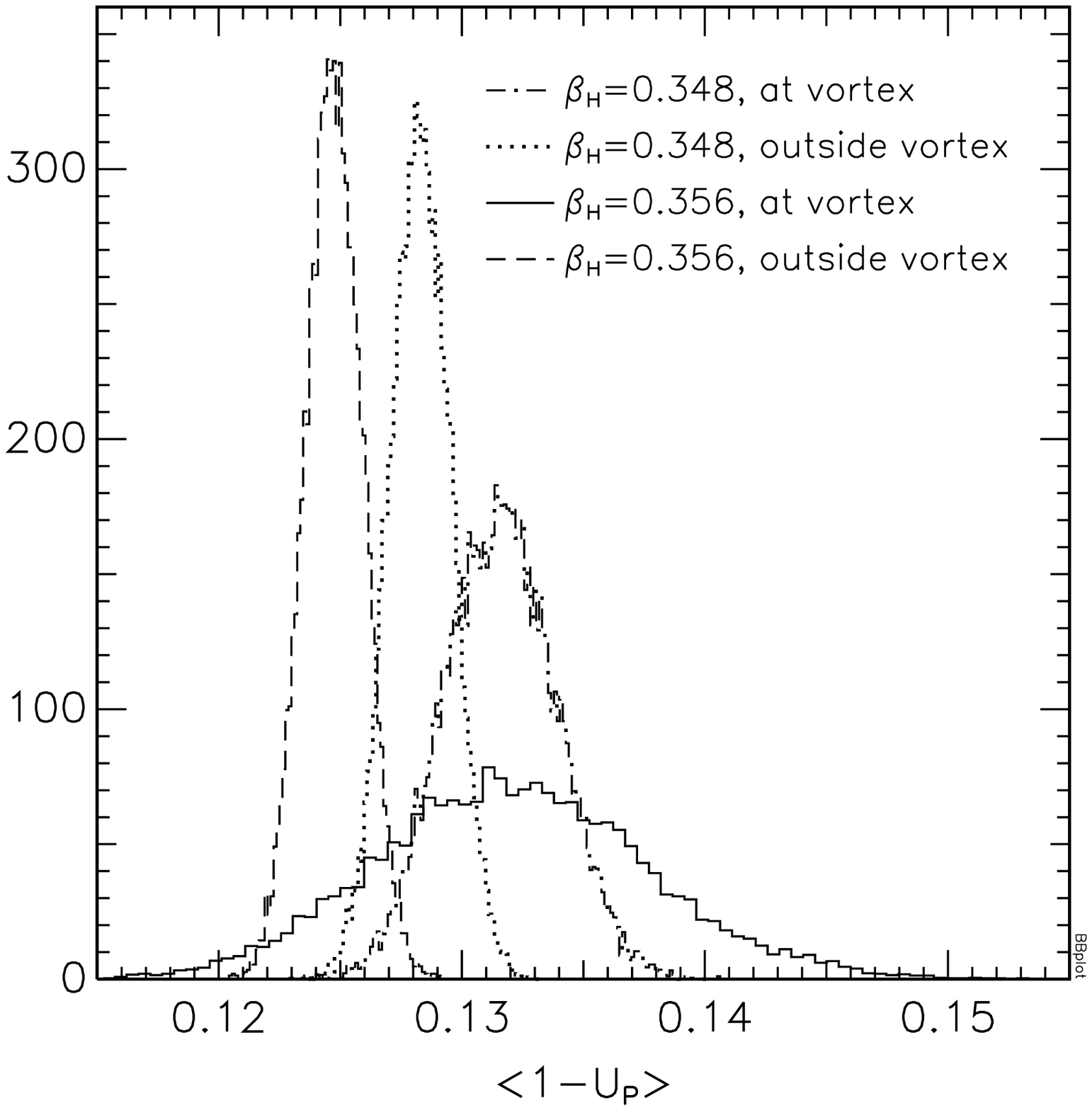} \\
(c) & \hspace{1.5cm}  (d) \\
\end{tabular}
\end{center}
\vspace{-0.5cm}
\caption{
The squared Higgs modulus inside and outside
of a $Z$--vortex: (a) {\it vs.} $\beta_H$; (b) histograms
on the symmetric ($\beta_H=0.348$) and on the Higgs ($\beta_H=0.356$)
sides of the percolation transition. The same for gauge field energy:
Figs. (c) and (d) respectively.}
\label{corr}
\end{figure*}

\end{document}